\documentclass[12pt]{iopart}
\usepackage{graphicx}

\newcommand{\gtrsim}
{\mbox{\raisebox{-0.7ex}{$\textstyle \sim$}
 \raisebox{ 0.3ex}{$\textstyle  \!\!\!\!\!\! >$  }}}

\newcommand{\lesssim}
{\mbox{\raisebox{-0.7ex}{$\textstyle \sim$}
 \raisebox{ 0.3ex}{$\textstyle  \!\!\!\!\!\! <$  }}}
\begin{document}
\title[Nonlinear evolution of dark matter and dark energy]{Nonlinear
 evolution of
 dark matter and dark energy in the Chaplygin-gas cosmology
 }
\author{Neven Bili\'{c}\dag ,
Robert J Lindebaum\ddag ,
Gary B Tupper\S , and Raoul D Viollier\S}
 \address{\dag\
 Rudjer Bo\v skovi\'c Institute,
P.O. Box 180, 10002 Zagreb, Croatia}
\address{\ddag
School of Chemical and Physical Sciences,
University of Natal,
Private Bag X01,
Scottsville 3209, South Africa}
\address{\S
Institute of Theoretical Physics and Astrophysics,
 Department of Physics, University of Cape Town,
 Private Bag, Rondebosch 7701, South Africa}

 \ead{
 \mailto{bilic@thphys.irb.hr},
\mailto{lindebaumr@nu.ac.za},
 \mailto{viollier@physci.uct.ac.za}}

\begin{abstract}
The hypothesis that dark matter and dark energy are
unified through the Chaplygin gas,
an exotic fluid obeying $p = - A/\rho$, is reexamined.
Using generalizations of the spherical model which incorporate
effects of the acoustic horizon we show that an initially
perturbative Chaplygin gas evolves into a mixed system
containing cold dark matter like gravitational condensate.
\end{abstract}

\maketitle

\newpage


\section{Introduction}
It has long been appreciated that there exists a dark-matter
problem in that the nonrelativistic matter fraction $\Omega_{\rm M}$ of critical
density, inferred from galactic rotation curves,
cluster stability, and peculiar
velocities, far exceeds the ordinary baryonic matter fraction
$\Omega_{B} \simeq 0.04$
 provided by nucleosynthesis, but falls
short of the standard inflationary
prediction of unity \cite{turn1}.
 Observations of
 high-redshift supernova \cite{perl2} and microwave background
anisotropies \cite{hal3,ben4} have now settled the case for a flat
universe with the dominant
component
dubbed `dark energy'
possessing appreciable negative pressure,
such that the ratio
 $w\equiv p/\rho$ is less than $-0.8$.
The most conservative reading of the WMAP results
\cite{ben4} is that $\Omega_{\rm CDM} \simeq 0.23$ is
provided by traditional cold dark matter (CDM) candidates,
while the balance resides in a
cosmological constant $\Lambda$
with  the corresponding density fraction
 $\Omega_{\Lambda} \simeq 0.73$.
The inherent fine-tuning and coincidence problems
of the $\Lambda$CDM model are somewhat ameliorated in
quintessence models \cite{peeb5},
 which replace $\Lambda$ by an evolving scalar field.
However,
like its predecessor,
a quintessence-CDM model
assumes that dark matter and dark
energy are distinct entities.
For a recent review on the most popular dark-matter and dark-energy models,
see \cite{sah}.

A more liberal interpretation is that these data may tell us something
profound about the space-time we inhabit,
dark matter/energy being different
manifestations of a common structure encoded in the dark sector's $w$ \cite{hu44}.
An interesting possibility
becomes evident upon recalling that the
average density in a virialized halo is more than several hundred
times higher than the critical density, suggesting
a density dependent $w=w(\rho)$ with
$w \simeq 0$ for
$\rho \gg \rho_{\rm cr}$,
$w  \sim -1$ for
$\rho \sim \rho_{\rm cr}$
and the identification of
``dark matter'' with objects which became nonlinear at high redshift.
The general class of models  in which a unification of dark matter and dark energy
is achieved through a single entity is often referred to as
{\em quartessence} \cite{mak1,mak2}.
Among other scenarios of unification that have recently been suggested,
interesting attempts are based on
the so-called {\em k-essence} \cite{chi,sch}, a scalar field with
noncanonical kinetic terms which was first introduced as a model for
inflation \cite{arm}.

Perhaps the simplest scenario in which dark matter and dark energy are
different manifestations of a single substance
may be realized through the Chaplygin gas,
an exotic fluid obeying
\begin{equation}
p =  - \frac{A}{\rho} \; ,
\label{eq001}
\end{equation}
which has been extensively studied for its
mathematical properties \cite{jack8}.
The cosmological potential of equation (\ref{eq001}) was first
noted by Kamenshchik {\it et al} \cite{kam9} who observed that
integrating the energy
conservation equation in a homogeneous model led to
\begin{equation}
\rho(a) = \sqrt{A + \frac{B}{a^{6}} }  \; ,
\label{eq002}
\end{equation}
where $a$ is the scale factor normalized to
unity today and $B$ an integration constant.
Thus, the Chaplygin gas interpolates between matter,
 $\rho \sim a^{-3}$,
at high redshift and a cosmological constant like
$\rho \sim \sqrt{A}$ as $a$ tends to infinity.

Of particular interest is that
equation (\ref{eq001}) is obtained \cite{jack8,bil6,bil7} from
the Lagrangian
\begin{equation}
{\cal L}  =
 - \sqrt{A} \; \sqrt{1 - g^{\mu \nu}
 \, \theta,_{\mu} \, \theta,_{\nu} } \; ,
\label{eq003}
\end{equation}
which may be seen by evaluating the stress-energy tensor $T_{\mu \nu}$ and  introducing
$u_{\mu} = \theta,_{\mu} / \sqrt{g^{\alpha \beta} \,
\theta,_{\alpha} \, \theta,_{\beta} }$
for the four-velocity and
$\rho = \sqrt{A} / \sqrt{1 - g^{\mu \nu} \,
\theta,_{\mu} \, \theta,_{\nu} }$
for the energy density. One recognizes $\cal{L}$
as a  Lagrangian of the Born-Infeld type familiar
in the $D$-brane constructions of string/$M$ theory \cite{jack8}.
Geometrically, $\cal{L}$
describes space-time as a 3+1 brane in a 4+1 bulk via
the embedding coordinate $X^{5} = \theta$.
The same Lagrangian appears
as the leading term in Sundrum's \cite{sun10} effective
field theory approach to large extra dimensions; for one such
extra dimension, $0 \leq \theta \leq \ell_{5}$, lowering the
Planck scale to the intermediate scale
of about
 10$^{11}$ GeV, one estimates
$A^{1/8} \sim \ell_{5}^{-1}$ $\sim$ meV \cite{bil7}.

 The Born-Infeld Lagrangian (\ref{eq003}) is a special
 case of the string-theory inspired
 tachyon Lagrangian \cite{sen} in which
the constant $\sqrt{A}$ is replaced by a potential $V(\theta)$.
Unfortunately, tachyon models encounter the problem that,
generically, $V(0)$ is a maximum, so the tachyon
field rolls quickly towards the minimum $V (\infty) = 0$ thereby
driving $w = \dot{\theta}^{2} - 1 \rightarrow 0$ to leave
only a dust component \cite{gibb12}.
However, the situation may be quite different in
inhomogeneous tachyon cosmology \cite{pad}.

It is possible to define a `generalized Chaplygin gas' \cite{bent11}
through
\begin{equation}
p = - A/\rho^{\alpha},
\label{eq104}
\end{equation}
where $\alpha \geq 0$.
In this case, the analogue
of equation (\ref{eq003}) lacks any
geometrical interpretation, 
but equation (\ref{eq104}) can be obtained from a moving brane in Schwarzschild-anti-de-Sitter bulk \cite{nev45}.
In the limit $\alpha\rightarrow 0$, one approaches a model
equivalent to $\Lambda$CDM in the background equations, although
the perturbation equations yield
a different power spectrum \cite{fab2}.


All models that unify dark matter and dark energy face the problem of
nonvanishing sound speed and the well-known Jeans instability.
A fluid with a nonzero sound speed has a characteristic scale below which
the pressure effectively opposes gravity. Hence the perturbations of the scale
smaller than the sonic horizon will be prevented from growing.
In the Chaplygin-gas model, the sound speed is small at early times
but  is significant at the crossover region and hence
the small scale perturbations are dumped. Indeed, it has been demonstrated
that the  purely  perturbative Chaplygin-gas model \cite{cun42,mak43} is in severe conflict
with the cosmic microwave background (CMB) angular power spectrum
\cite{bent2,cart23} and the mass power spectrum \cite{sand18}.
However, it has recently been pointed out that if nonadiabatic
perturbations are allowed, the Chaplygin-gas model may be compatible
with the 2df mass power spectrum \cite{rei}.

A  more general situation where
the Chaplygin gas is mixed with CDM
is considered in a
 number of papers \cite{dev38,dev39,avel40,alca41,bean19,ame,mult,bert}.
In this case, the Chaplygin gas simply plays the role of  dark energy.
In keeping with the quartessence
philosophy it would be preferred if  cold dark matter could be replaced by
droplets of Chaplygin condensate.
Homogeneous world models containing a mixture of cold dark matter and
Chaplygin gas have been successfully confronted with lensing statistics
\cite{dev38,dev39}
as well as with supernova and other tests
\cite{avel40,alca41}.
Bean and Dor\'{e} \cite{bean19} and similarly Amendola {\it et al}
\cite{ame} have
examined a mixture of CDM and the generalized Chaplygin
gas against supernova, large-scale structure, and CMB constraints
and demonstrated that a
thorough likelihood analysis favors the limit $\alpha
\rightarrow 0$, i.e., the equivalent $\Lambda$CDM
model.  In both papers it has been concluded that the standard Chaplygin gas
is ruled out as a candidate for dark energy.
However, a recent analysis \cite{bert,bert2} of the supernova data
seems to indicate that the generalized Chaplygin gas with
$\alpha\geq 1$ is favored over the $\alpha\rightarrow 0$ model.

In spite of these somewhat discouraging results
we share the opinion of Gorini {\it et al}
\cite{gor1} that
the Chaplygin gas  still deserves further investigation
and that
a thorough investigation of the nonlinear regime
of the growth of inhomogeneities is needed for a definite
conclusion concerning the compatibility of the Chaplygin gas cosmologies
with the observable large-scale structure of the universe.
On the basis of an extended spherical model
which includes the sonic horizon
effects,
in this paper we
show
 that a
fraction of the Chaplygin gas
condensates and never reaches a stage where its properties change from
dark-matter-like to dark-energy-like.
Hence,  at the crossover region a fraction of the gas behaves effectively as dark matter
and the rest as dark energy.

We organize the paper as follows. In section 2 we discuss
the nonlinear evolution of  perturbation in the Newtonian limit
by generalizing the spherical model.
Summary and conclusions are given in section 3.
\section{Condensate Formation}
To be able to claim that the Chaplygin gas (or any other candidate)
actually achieves unification, one must be assured that
initial perturbations can evolve into a deeply nonlinear regime
to form a gravitational condensate of superparticles that
can play the role of cold dark matter. In \cite{bil6, bil7} this was
inferred on the basis of the Zel'dovich
approximation \cite{zel13}.
In comoving coordinates,
the solution for the inhomogeneous Chaplygin-gas cosmology
is
\begin{equation}
\rho = \sqrt{A + \frac{B}{\gamma}}.
\label{eq13}
\end{equation}
Here $\gamma$ is the determinant of the induced spatial metric tensor
$\gamma_{ij} = g_{i0} \, g_{j0}/g_{00} - g_{ij}$
 and $B$ can be taken as constant on the
scales of interest.
The generalization
(\ref{eq13}) of equation (\ref{eq002}) allows us to implement the
geometric version
of the Zel'dovich approximation:
the transformation from Euler to Lagrange (comoving)
coordinates induces $\gamma_{ij}$ as
\begin{equation}
\gamma_{ij} = \delta_{kl}  {D_{i}}^{k}  {D_{j}}^l\, , \hspace{1cm}
{D_{i}}^{j} = a({\delta_{i}}^{j}-b{\varphi_{,i}}^{j}) \, ,
\label{eq14}
\end{equation}
where
${D_{i}}^{j}$ is the deformation tensor,
$\varphi$ is the velocity potential, and  the
quantity $b=b(t)$ describes the evolution of the perturbation.
The Zel'dovich approximation offers a means
of extrapolation into the
nonperturbative regime via equation (\ref{eq13}) and
\begin{equation}
\gamma = a^{6}
(1-\lambda_{1}b)^2 (1-\lambda_{2}b)^2 (1-\lambda_{3}b)^2,
\label{eq17}
\end{equation}
where the $\lambda_{i}$ are the eigenvalues of
${\varphi_{,i}}^{j}$.
When one
(or more) of the $\lambda$'s is (are) positive,
a caustic forms on which $\gamma
\rightarrow 0$ and $w \rightarrow 0$, i.e.,
at the locations where structure
forms the Chaplygin gas behaves as dark matter.
Conversely, when all of the
$\lambda$'s are negative, a void forms,
$\rho$ is driven to its limiting value
$\sqrt{A}$, and the Chaplygin gas
behaves as dark energy, driving accelerated
expansion.

For the issue at hand, the 
usual
Zel'dovich approximation has the
shortcoming that the effects of finite sound speed are neglected.
Since the structure formation occurs in the decelerating phase,
we can address the question in the Newtonian limit
$p\ll\rho$
by generalizing the spherical model.
In the case of vanishing shear and rotation,
the continuity and Euler-Poisson equations become
\begin{equation}
\dot{\rho} + 3 {\cal{H}} \, \rho  =  0 ,
\label{eq007a}
\end{equation}
\begin{equation}
3 \dot{\cal{H}} + 3 {\cal{H}}^{2} +
4 \pi G \rho + \frac{\partial}{\partial r_i}
\left( \frac{c_s^{2}}{\rho}
\frac{\partial}{\partial r_i}
  \rho \right)  =  0 ,
\label{eq007b}
\end{equation}
where $c_s$ is the adiabatic speed of sound given by
\begin{equation}
c_s^2=\frac{\partial p}{\partial\rho}=\frac{A}{\rho^2},
\label{eq407}
\end{equation}
$\cal{H}$ is the local Hubble parameter, and the time derivative
is at the fixed Lagrangian coordinate $\vec{r}$.
It is convenient to introduce the comoving coordinates
$\vec{x}=\vec{r}/a$.
Then writing
\begin{equation}
\rho =\bar{\rho}(1+\delta) ,
\label{eq107}
\end{equation}
\begin{equation}
{\cal H}= H +\delta{\cal{H}} ,
\label{eq207}
\end{equation}
subtracting the background,
eliminating $\delta{\cal H}$ and
 $\dot{\delta{\cal H}}$,
and changing the variables from $t$ to $a$
we find
\begin{equation}
a^{2}  \delta'' + \frac{3}{2}\, a \delta' -
\frac{3}{2}\, \delta (1 + \delta)-\frac{4}{3}\,
\frac{(a \delta')^2}{1 + \delta} -
\frac{1+\delta}{a^2 H^2}
\frac{\partial}{\partial x_i}
\left( \frac{c_s^2}{1+\delta}
\frac{\partial\delta}{\partial x_i}\right)=0 ,
\label{eq108}
\end{equation}
where $'$ denotes the derivative with respect to $a$.
To solve this partial differential equation as it stands, one would need a rather
involved numerical computation. Instead, we proceed first by
discussing the linear approximation of (\ref{eq108}) and then by solving
the nonlinear problem imposing
certain restrictions on the solution.

First, consider the linear approximation which
has been discussed in detail by Fabris {\it et al} \cite{fab14}.
Keeping only the terms linear in $\delta$, $\delta'$, and $\delta''$
and using
the expansion
\begin{equation}
\delta(a,\vec{x})=\sum_k \delta_{\rm pert}(k,a) e^{i\vec{k} \vec{x}},
\label{eq208}
\end{equation}
 one obtains an explicit solution for the perturbative density contrast
  which may be expressed as
\begin{equation}
\delta_{\rm pert} (k, a) \; \propto \;
a^{-1/4} J_{5/14}  (d_{\rm{s}} k) \, .
\label{eq004}
\end{equation}
Here $J_{\nu} (z)$ is the Bessel function, $k$ the comoving wave
number,
and $d_{\rm s}$ the comoving sonic horizon  given by
\begin{equation}
d_{\rm{s}}  =
\int_0^a \frac{\bar{c}_s da}{a^2 H} =
\frac{2}{7}
\frac{\left( 1 - \Omega^{2} \right)^{1/2}}{\Omega^{3/2}}
\frac{a^{7/2}}{H_{0}} \, ,
\label{eq005}
\end{equation}
with $\bar{c}_s=\sqrt{A}/\bar{\rho}$ and the equivalent matter fraction
\begin{equation}
\Omega = \sqrt{B/(A+B)} =
\sqrt{B}/\rho_{\rm cr}\, .
\label{eq105}
\end{equation}
Thus, for $ d_{\rm{s}}k \ll 1$, $\delta_{\rm pert} \sim a$,
but for $ d_{\rm{s}}k \gg 1$,
$\delta_{\rm pert}$
undergoes damped oscillations.
Similar results
have been obtained
numerically in the relativistic case \cite{fab15}.

Next, consider the nonlinear evolution.
To simplify equation (\ref{eq108}), it seems reasonable
to make the following restriction: we take
$\delta (t,\vec{x})$
to be of fixed shape, wave number dependent size,
    and time dependent amplitude, then evaluate
    the comoving coordinate derivatives at the
    origin. That is to say, we retain the spirit of
    the spherical dust model where each comoving
    coordinate point is treated as an independent
    origin, but allow for the local coupling of the
    inhomogeneity curvature to sound waves.
    A convenient choice is the spherical lump
\begin{equation}
    \delta (a,\vec{x}) = \delta_{R}(a) f (x/R) ,
\label{eq308}
\end{equation}
where $f(z)$ is an arbitrary function satisfying
$f'(0)=1$ and $f''(0) < 0$.
For simplicity, we take
 $f(z) = {\rm exp} (- z^{2}/2)$.
Evaluating (\ref{eq108}) at the origin, we obtain an
ordinary differential equation for $\delta_R$
\begin{equation}
a^{2}  \delta_{R}^{''} + \frac{3}{2}\, a \delta_{R}^{'} -
\frac{3}{2}\,\delta_{R} (1 + \delta_{R})-\frac{4}{3}\,
\frac{(a \delta_{R}^{'})^{2}}{1 + \delta_{R}} + \frac{49}{4} \,
 \left( \frac{a}{a_{R}} \right)^{7}
\frac{\delta_{R}}{(1 + \delta_{R})^{2}}  = 0  ,
\label{eq008}
\end{equation}
where
\begin{equation}
a_{R}^{-1}\equiv (d_{\rm{s}} \sqrt{3}/R )^{2/7} a^{-1} =
8.06\, \left( 1 - \Omega^{2} \right)^{1/7}
\Omega^{- 3/7} \left( R^{-1} h^{-1} \, {\rm Mpc} \right)^{2/7} .
\label{eq307}
\end{equation}
The recommendations of equation (\ref{eq008}) are that it
\begin{itemize}
\item[(i)]  reproduces equation (\ref{eq004}) at linear order and
\item[(ii)] extends the spherical dust model by
incorporating the Jeans length through the last term.
\end{itemize}

To gain some perspective,
with WMAP fixing $a_{\rm eq} = 3 \cdot 10^{-4}$
and reionization at a redshift of 20, initial perturbations must
satisfy $\delta (a_{\rm eq}) \; \gtrsim \; 6 \cdot 10^{-3}$
on mass scales $M \lesssim 10^{10} M_{\odot}$ even for
$\Lambda$CDM.
Taking $\Omega=0.27$ and $h=0.71$, we find
 $a_{R}^{-1}=15.4 ( R^{-1} \,\mbox{Mpc})^{2/7}$,
which coincides with $1+z_{\rm reion}=21$
at $R$ = 0.338 Mpc
corresponding to a protogalaxy mass
$M$ = 2.3
$\cdot 10^{10} M_{\odot}$.
\begin{figure}
\begin{center}
\includegraphics[width=.5\textwidth,trim= 0 2cm 0 2cm]{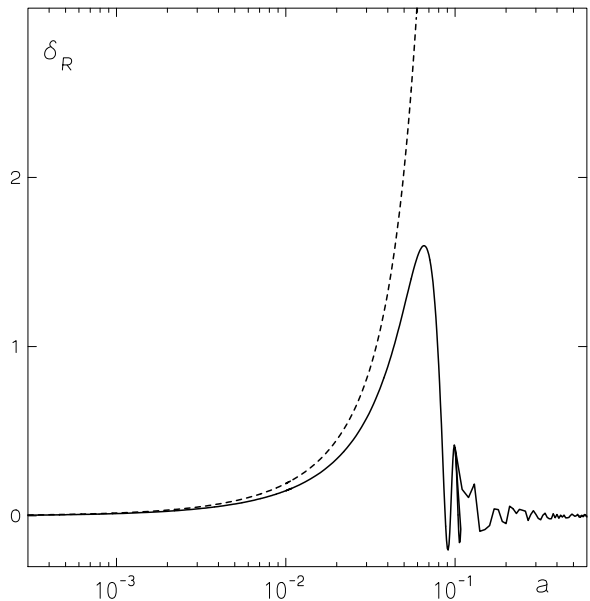}
\caption{
 Evolution of $\delta_{R}(a)$ in
            the spherical model, equation (\ref{eq008}),
           from $a_{\rm eq} = 3\times 10^{-4}$
           for $a_{R}$ = 0.0476,
           $\delta_{R} (a_{\rm eq})$ =0.004 (solid)
            and $\delta_{R} (a_{\rm eq})$ =0.005 (dashed).
}\label{fig1}
\end{center}
\end{figure}

\begin{figure}
\begin{center}
\includegraphics[width=.5\textwidth,trim= 0 2cm 0 2cm]{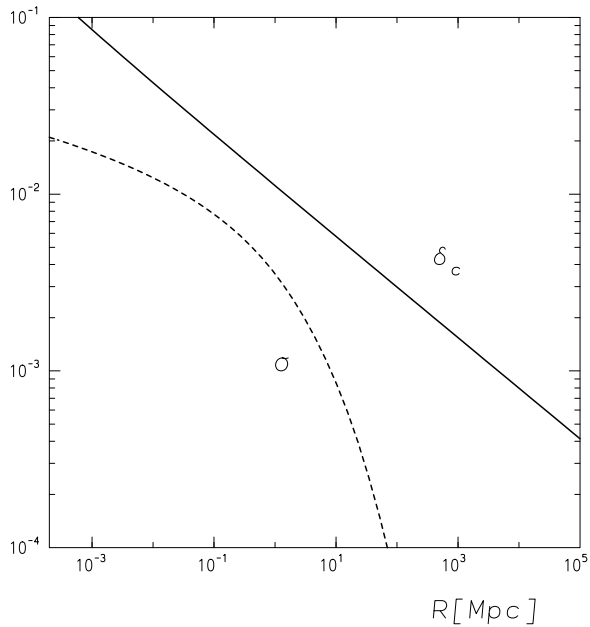}
\caption{
Initial value $\delta_{R}(a_{\rm dec})$  versus $R$
for $\Omega=0.27$ and $h=0.71$.
The threshold $\delta_c (a_{\rm dec})$
is shown by the line separating the two regimes.
The dashed line gives $\sigma (R)$ calculated using the
concordance model.
 }\label{fig2}
\end{center}
\end{figure}

In figure \ref{fig1} we show the evolution of two initial
perturbations from the radiation-matter equality
for $a_{R}$ = 1/21.
In contrast to linear theory, where for any $R$ the acoustic
horizon will eventually stop $\delta_R$ from growing irrespective of the initial
value of the perturbation,
here, for initial $\delta_{R} (a_{\rm dec})$ above a certain
threshold, $\delta_{R} (a) \rightarrow \infty$ at
finite $a$ just as in the dust model.
Conversely, at sufficiently small
$\delta_{R} (a_{\rm dec})$, the acoustic horizon can stop
$\delta_{R} (a)$ from growing even in a mildly nonlinear regime.
Figure \ref{fig2} shows how the threshold
$\delta_R(a_{\rm dec})$ divides the two regimes
depending on the comoving scale $R$.
Qualitatively similar conclusions have been reached
in a different way
by Avelino {\it et al}
\cite{avel35}.

One question is how accurate is the approximation (\ref{eq308}) which
assumes that the comoving size is time independent?
To check this, we have relaxed the assumption (\ref{eq308})
by allowing the second derivative of $f$ at the origin to
depend on $a$, i.e., the initial Gaussian perturbation
to have a variable width.
Evolution of the width is then described by a second-order differential
equation coupled to  equation (\ref{eq008}).
By solving this system of coupled equations
we find a slight change of the behavior of
$\delta_R$ in the linear regime but no significant change
of the linear/nonlinear boundary depicted in figure \ref{fig2}.

\begin{figure}
\begin{center}
\includegraphics[width=.5\textwidth,trim= 0 2cm 0 2cm]{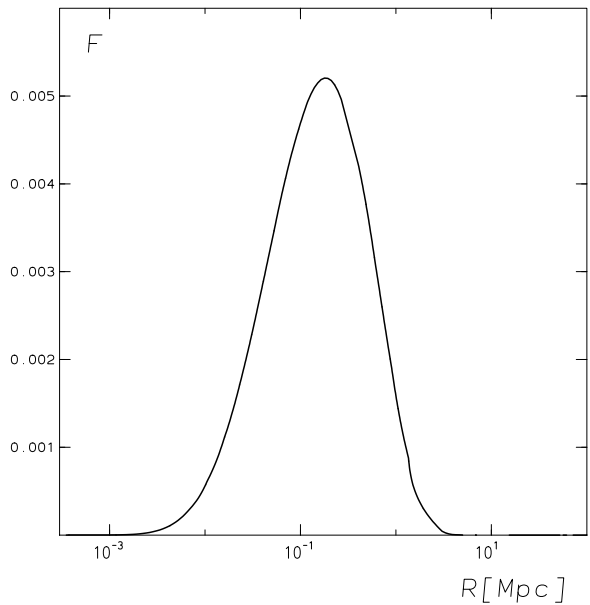}
\caption{
Fraction of Chaplygin gas in collapsed objects using
$\delta_c(R)$ and $\sigma(R)$ from
figure \ref{fig2}.
 }\label{fig3}
\end{center}
\end{figure}

The crucial question is what fraction of Chaplygin gas goes into condensate.
Indeed, in \cite{bil36} we have noted that if this were 92\% as given by the
geometric Zel'dovich approximation, the CMB and the mass power spectrum
would be reproduced.
To answer this quantitatively, we adopt the
 Press-Schechter
approach \cite{pres37}.
Assuming $\delta_{R}(a_{\rm dec})$ is given by a Gaussian random field with
dispersion $\sigma(R)$, and
including the notorious factor of 2 to
account for the cloud in cloud problem,
the condensate fraction at
a scale $R$ is given by
\begin{equation}
F (R) = 2 \int_{\delta_{c}(R)}^{\infty} \; \frac{d \delta}{ \sqrt{2 \pi} \sigma (R) }\;
{\rm exp} \left( - \frac{\delta^{2}}{2 \sigma^{2}(R)} \right)
= {\rm erfc} \left(
\frac{ \delta_{c} (R) }{ \sqrt{2} \; \sigma (R) }
\right) \, ,
\label{eq401}
\end{equation}
where $\delta_c(R)$ is the threshold $\delta_c(a_{\rm dec})$
depicted in figure \ref{fig2}. In figure \ref{fig2} we also exhibit
the dispersion
\begin{equation}
\sigma^{2}(R) = \int_{0}^{\infty} \; \frac{dk}{k} \; {\rm exp}
( - k^{2} R^{2} ) \Delta^{2} (k, a_{\rm dec}) ,
\label{eq402}
\end{equation}
calculated using $f$ above as the window function  and the concordance
model \cite{ben4} variance
\begin{equation}
\Delta^{2} (k,a)={\rm const} \left(\frac{k}{a H}\right)^4
T^2(k)\left(\frac{k}{7.5 a_0 H_0}\right)^{n-1}\, .
\label{eq403}
\end{equation}
In figure \ref{fig3} we present $F(R)$ calculated
using (\ref{eq401})-(\ref{eq403}) with
const=7.11$\times 10^{-9}$, the spectral index $n$=1.02,
and the parametrization of Bardeen {\it et al} \cite{bar} for the transfer
function $T(k)$ with $\Omega_{\rm B}$=0.04.
The parameters are fixed by fitting (\ref{eq403})
to the 2dFGRS power spectrum data \cite{perc}.
Our result, in qualitative agreement with
that of Avelino et al
\cite{avel35},
demonstrates that the collapse fraction is less than 1\% ,
far too small to affect the conclusions of
\cite{cart23,sand18}.

\section{Summary and Conclusions}
Thus we have found that nonlinear condensate, while present, is insufficient to
save the simple Chaplygin gas. In effect, the model is a victim of the
radiation dominated phase which turns the Harrison-Zel'dovich spectrum
$\delta_{k} \sim k^{1/2}$ to
$\delta_{k} \sim k^{- 3/2}$ at
$R_{\rm eq} \simeq {\rm 26 Mpc}$.
In the pure Chaplygin-gas universe there would inevitably be insufficient
small scale power to drive condensation.
This is a more
definitive, if pessimistic, statement than [43] because we have actually
followed the perturbations into the fully nonlinear regime.

Pragmatically, one can follow
\cite{hu44}
by introducing entropy perturbations to make the effective speed of
sound vanish, even in the nonperturbative regime
\cite{rei}.
Caution should, however, be taken that imposing $\delta p = 0$
as an initial condition does not guarantee its propagation except
in the long-wavelength limit
\cite{abra46}
where there is no problem in any case. More important is that
the negative results were drawn without exploiting the braneworld
connection of
equation (\ref{eq003}):
in braneworld models (for a recent review, see
\cite{maar47})
the Einstein equations are modified, e.g., by dark radiation and
hence so are
equations (\ref{eq007a},\ref{eq007b}).
From a different perspective, similar changes are brought about
by the radion mode
\cite{kim48}
which gives a scalar-tensor gravity.

An interesting idea how to circumvent the difficulties with structure
formation in the Chaplygin gas model has recently been put forward
by Bento {\it et al} \cite{bennew}.
Following \cite{pad}, they have identified $p = -\rho_{\Lambda}$ and $\rho_{\rm dm} = \rho+p$, so that the
energy-momentum tensor appears as a combination of dark matter $\rho_{\rm dm}$
and the cosmological ``constant'' $\Lambda$.
Still, $T^{\mu}_{\nu ; \mu} = 0$ supplies
the continuity and Euler equations while $\rho +3p = \rho_{\rm dm}
 -2\rho_{\Lambda}$
sources the gravitational potential. In particular,
$(\partial \rho_{\Lambda}/ \partial r_i)/\rho_{\rm dm}$ remains in the Euler equation.
Since $|p| \ll \rho \simeq \rho_{\rm dm}$ in the Newtonian context, and
$c_{\rm dm}^{2} =- \partial\rho_{\Lambda}/ \partial \rho_{\rm dm}$ =
$c_{s}^{2} / \left( 1 + c_{s}^{2} \right) \simeq c_{s}^{2}$,
there is no modification to equations (\ref{eq007a})
and (\ref{eq007b}), or the conclusions on
structure formation. The reason Bento et al \cite{bennew} conclude this
bookkeeping allows $\delta_{\rm dm}$ to grow unimpeded is that
they have taken over the equations of Arcuri and Waga \cite{arc},
wherein  $\Lambda$ is assumed
time dependent but homogeneous - {\it cf} equation (14) therein. Here,
instead, $\Lambda$ homogeneous implies $\rho_{\rm dm}$ homogeneous,
i.e. $\delta_{\rm dm}$ = 0.

Our prognosis is then that the crisis in quartessence cosmology
is akin to that which existed in electroweak gauge theory before the
Higgs mechanism. The Chaplygin-gas model is not so much wrong as
incomplete.

\section*{Acknowledgments}
Two of us, GBT and RDV wish to thank Jihn Kim for many
enlightening discussions on the braneworld aspects of the
Chaplygin gas.
This research is in part supported by
the Foundation for Fundamental Research
 (FFR) grant number  PHY99-1241 and the Research Committee of the
 University of Cape Town.
The work of NB is supported
 in part by the Ministry of Science and Technology of the
 Republic of Croatia under Contract No. 0098002.

\end{document}